\documentstyle [twoside,12pt]{article}
\newcommand{\nc}{\newcommand}
\nc{\la}{\lambda} \nc{\al}{\alpha}
\nc{\th}{\theta}  \nc{\be}{\beta}
\nc{\ga}{\gamma}  \nc{\Ga}{\Gamma}
\nc{\de}{\delta}  \nc{\De}{\Delta}
\nc{\si}{\sigma}  \nc{\ka}{\kappa}
\nc{\om}{\omega}  \nc{\Om}{\Omega}
\nc{\nf}{\infty}
\nc{\ra}{\rightarrow}
\nc{\beq}{\begin{equation}}
\nc{\eeq}{\end{equation}}
\nc{\beqa}{\begin{eqnarray}}  \nc{\dst}{\displaystyle}
\nc{\eeqa}{\end{eqnarray}} \nc{\nnb}{\nonumber}
\title{ \bf Local heterotic geometry in holomorphic coordinates }
\author{Guy Bonneau\thanks
{\noindent Laboratoire de Physique Th\'eorique et des Hautes Energies,
 Unit\'e associ\'ee au CNRS URA 280,~Universit\'e Paris 7,
 2 Place Jussieu, 75251 Paris Cedex 05.}
\and
Galliano Valent$^{\ast}$}
\date{ }
\begin{document}
\maketitle
\begin{abstract}
\noindent In the same spirit as done for N=2 and N=4 supersymmetric non-linear
$\si$ models in 2 space-time dimensions by Zumino and Alvarez- Gaum\'e and
Freedman, we analyse the (2,0) and (4,0) heterotic geometry in holomorphic
coordinates. We study the properties of the torsion tensor and give the
conditions under which (2,0) geometry is conformally equivalent to a (2,2) one.
Using additional isometries, we show that it is difficult to equip a manifold
with a closed torsion tensor, but for the real 4 dimensional case where we
exhibit new examples.
We show that, contrarily to Callan, Harvey and Strominger 's claim for real 4
dimensional manifolds, (4,0) heterotic geometry is not necessarily conformally
equivalent to a (4,4) K\"ahler Ricci flat geometry. We rather prove that,
whatever the real dimension be, they are special quasi Ricci flat spaces, and
we exemplify our results on Eguchi-Hanson and Taub-NUT metrics with torsion.
\end{abstract}

\vfill
{\bf PAR/LPTHE/93-56}\hfill  December 1993

\section{Introduction}

In 1979, B. Zumino \cite{[1]} proved the importance of complex geometry for the
study of supersymmetric $\sigma$-models. In particular, using holomorphic
coordinates, he showed that \newline N = 2 supersymmetry in two space-time
dimensions requires the bosonic part of the Lagrangian density to be built on a
K\"ahler manifold.
\noindent The work of Alvarez-Gaum\'e and Freedman  has generalised  this
approach to N = 4 supersymmetry which was shown to require hyperk\"ahler
manifolds \cite{[2]}. The heterotic supersymmetries lead to generalisations of
the aforementioned geometries : in the (2,0) \cite{[33],[35]} and in the (4,0)
cases \cite{[23],[24],[77],[3]}, it was shown that one needs target manifolds
with torsion. It is the aim of the present work to analyse the properties of
such manifolds equipped with torsion using holomorphic coordinates.

In section 2, we recall the necessary and sufficient conditions on the target
space metric and torsion tensors for (2,0) heterotic supersymmetry and express
them in holomorphic coordinates. Various geometrical objects are then given in
such coordinates, in the same spirit as done by Alvarez-Gaum\'e and Freedman
\cite{[4]} and Hull \cite{[35]}. A torsion potential $b_{\mu\nu}$ and a vector
$V_{\mu}$ (dual of the torsion tensor in the special case of 4 real dimensions)
are then introduced and used to express the relevant geometrical objects.

\noindent We then analyse the necessary and sufficient conditions under which,
given a complex Riemannian manifold of real dimension 2N, a (2,0) heterotic
geometry (metric $g_{\mu\nu}$ + torsion) and a (2,2) one (K\"ahler metric
$\hat{g}_{\mu\nu}$, no torsion) have conformally related metrics :
$$\hat{g}_{\mu\nu} = e^{-2f}g_{\mu\nu} \ .$$
This appears to be a very restrictive condition as the conformal factor has to
be a real function, harmonic with respect to the K\"ahler metric laplacian,
whatever the dimension of the manifold be: $$\hat{\De}e^{2f} = 0\ .$$

\noindent In order to describe geometries leading to one-loop finite non-linear
$\si$ models in two space time dimensions, following Friedan \cite{[8]} in the
torsionless case, Friedling and Van de Ven \cite{[9]} introduced  ``
generalised quasi Ricci flat" metrics through \footnote{
$A_{\lambda[\mu\,\nu]\rho}\stackrel{\rm def}{=}\
{1\over2}(A_{\lambda\mu\nu\rho}\ -\ A_{\lambda\nu\mu\rho})$ and
$A_{\lambda(\mu\,\nu)\rho}\stackrel{\rm def}{=}\
{1\over2}(A_{\lambda\mu\nu\rho}\ +\ A_{\lambda\nu\mu\rho})$.
Moreover $\nabla_{\mu}$ is the ordinary covariant derivative with the
symmetric Christoffel connection.\ }  :
\beqa\label{a1}
\exists\ \ W_{\mu}\ {\rm and}\ \chi_{\mu} \ \   {\rm such\ that} \ \
Ric_{(\mu\nu)} & = &  \nabla_{(\mu}W_{\nu)} \nnb \\
 Ric_{[\mu\nu]} & = & {1\over 2}T^{\rho}_{\mu\nu}W_{\rho} +
\nabla_{[\mu}\chi_{\nu]}\ \ .
\eeqa
We then express these conditions in holomorphic coordinates and also discuss
the conditions that restrict the holonomy group from U(N) to SU(N) : indeed, it
was argued by Hull \cite{[35]} that such a requirement is necessary for
off-shell one-loop finiteness of (2,0) non-linear $\si$ models in two space
time dimensions. We show that the metric has to be a special ``generalized
quasi Ricci flat" one \footnote{\ $\chi_{\mu} = - W_{\mu}$ where  $W_{\mu}$ is
related to the trace of the torsion tensor. }.

Several examples of metrics with special isometries are then constructed in
Section 3. Despite the large isometry groups considered, which enforce
conformal equivalence with the K\"ahler torsionless case, (2,0) supersymmetry
is not sufficient to define uniquely the metrics and several examples of
supplementary properties are analysed. In particular, we generalise the
(torsionless) metrics of LeBrun \cite{[44]} which have a vanishing scalar
curvature, to target spaces with torsion and obtain as a special case an easy
derivation of Eguchi-Hanson metric with torsion \footnote{\  first obtained by
Delduc and Valent \cite{[6]}\ .} directly in holomorphic coordinates.
Unfortunately, despite their conformal equivalence with regular K\"ahler ones,
all the metrics we obtained are singular.

Section 4 is devoted to (4,0) heterotic geometry. The necessary and sufficient
conditions for (4,0) supersymmetry are given
(as shown in \cite{[77]}, they are slightly less restrictive than often
asserted ( see for example \cite{[3]})), and we prove that in any 4N
dimensional case, the metric has to be a special `` generalised quasi Ricci
flat" one  where the vector $V_{\mu}$ depends on a single complex function F.
We also show  that scalar flatness is obtained on general grounds only for 4
real dimensions.

\noindent We then analyse the possible conformal equivalence of such metrics
with torsionless hyperk\"ahler ones involved in (4,4) supersymmetry and show
that, even in 4 real dimensions, contrarily to Callan et al.'s claim
\cite{[5]}, (4,0) world sheet supersymmetry does not imply that the
corresponding metric is conformally equivalent to a Ricci flat K\"ahler one.

Section 5 is then devoted to a detailed study of Eguchi-Hanson and Taub-NUT
metrics with torsion \cite{[6]} in holomorphic coordinates. While Eguchi-Hanson
with torsion is indeed conformally equivalent to its torsionless counterpart,
this fails to be true for Taub-NUT with torsion. To conclude some  remarks are
offered in Section 6.

\section{(2,0) heterotic geometry in complex coordinates}
\subsection{Generalities and notations }
As explained in the introduction, we consider 2N real dimensional complex
Riemannian manifolds with torsion and assume that the metric is hermitian with
respect to the covariantly constant complex structure J and that the torsion
tensor three form is closed. These conditions are necessary and sufficient to
build a (2,0) supersymmetric $\sigma$ model with a Wess-Zumino term
\cite{[33]}.
All these hypothesis write :

- distance : \beq \label{b1}
d\tau^{2}\ =\ g_{\mu\nu}dx^{\mu}dx^{\nu},\ \ \ \ g_{\mu\nu}\ =\ g_{\nu\mu}
\eeq

- torsion : the torsion tensor $T_{\mu\nu\rho} =
g_{\mu\sigma}T^{\sigma}_{\nu\rho}$ is fully skew-symmetric and its associated
three form is closed
\beq \label{b2}
T\ =\ \frac{1}{3!}T_{\mu\nu\rho}dx^{\mu}\wedge dx^{\nu}\wedge dx^{\rho},\ \ \
\ dT\ =\ 0
\eeq

- $J_{\mu}^{\nu}$ is an almost complex structure :
\beq\label{b3}
J_{\mu}^{\nu} J_{\nu}^{\rho} = -\ \delta_{\mu}^{\rho}
\eeq

- integrability condition on $J_{\mu}^{\nu}$ to be a complex structure :
\beq\label{b4}
N^{\rho}_{\mu\nu}\ \equiv \  J_{\mu}^{\la}(\partial_{\la} J_{\nu}^{\rho}\ -\
\partial_{\nu} J_{\la}^{\rho})\ -(\mu\ \leftrightarrow \ \nu) \ =\ 0
\eeq

- hermiticity : the metric is hermitian with respect to the complex structure
\beq\label{b7}
g_{\mu\la}J_{\nu}^{\la} \ + \ J_{\mu}^{\la}g_{\la\nu} \ =\ 0
\eeq
which is equivalent to the statement that $J_{\mu\nu}\ \stackrel{\rm def}{=}\
J_{\mu}^{\sigma}g_{\sigma\nu}$ is skew symmetric and therefore locally defines
a two-form :
\beq\label{b77}
\om = {1\over 2}J_{\mu\nu}dx^{\mu}\wedge dx^{\nu}
\eeq

- covariant constancy of $J_{\mu}^{\nu}$ :
\beq\label{b5}
D_{\mu}J_{\nu}^{\rho} \equiv \ \partial_{\mu}J_{\nu}^{\rho}
+\Gamma^{\rho}_{\la\mu} J_{\nu}^{\la} - \Gamma^{\la}_{\nu\mu} J_{\la}^{\rho} \
=\ 0
\eeq
(The connection with torsion is taken as
$\Gamma^{\rho}_{\mu\nu}=\gamma^{\rho}_{\mu\nu}-{1\over 2}T^{\rho}_{\mu\nu} $
where $\gamma^{\rho}_{\mu\nu}$ is the usual symmetric Christoffel connection).
As a consequence of equ. (\ref{b5}), the integrability condition (\ref{b4})
reduces to an algebraic constraint on the torsion :
\beq\label{b6}
N^{\rho}_{\mu\nu}\ \equiv \ T^{\rho}_{\mu\nu}\ -\
J_{\mu}^{\la}J_{\nu}^{\sigma}T^{\rho}_{\la\sigma} \ +\
(J_{\mu}^{\la}T^{\sigma}_{\la\nu}\ -\
J_{\nu}^{\la}T^{\sigma}_{\la\mu})J_{\sigma}^{\rho} \ =\ 0
\eeq
( compare with equ.(21) of \cite{[3]} ) and then one obtains :
\beq\label{b88}
d\om = -{1\over 2}T_{\mu '\nu \rho}J_{\mu}^{\mu '}dx^{\mu}\wedge dx^{\nu}\wedge
dx^{\rho} = {1\over 2}T_{\mu '\nu '\rho '}J_{\mu}^{\mu '}J_{\nu}^{\nu
'}J_{\rho}^{\rho '}dx^{\mu}\wedge dx^{\nu}\wedge dx^{\rho}
\eeq

\noindent In the absence of torsion, equations
(\ref{b1},\ref{b3},\ref{b7},\ref{b5}) define a K\"ahler manifold.

In the following, due to (\ref{b3},\ref{b4}), one can choose a coordinate
system where J is diagonal and constant. The real coordinates $x^{\mu}$ being
split into the complex ones $z^{i}$ and $ \bar{z}^{i}$, one has:
\beq\label{b8}
\left\{ J_{j}^{\bar{i}} =  J_{\bar{j}}^{i} = 0\ ,\hspace{0.5cm}
J_{j}^{i} = -J_{\bar{j}}^{\bar{i}}  = i\delta_{j}^{i}\ \right\}
\Leftrightarrow \ \left\{ J_{i\,j} = J_{\bar{i}\,\bar{j}} = 0\ ,\hspace{0.5cm}
J_{j\,\bar{i}} = -J_{\bar{i}\, j} = ig_{j\,\bar{i}}\ \right\}\ .
\eeq

\subsection{ The geometrical objects in complex coordinates}
In these complex coordinates, equation (\ref{b7}) gives :
$$g_{ij}\ =\ g_{\bar{i}\,\bar{j}}\ =\ 0 $$
The distance becomes :
\beq\label{b9}
d\tau ^{2}\ =\ 2g_{i\,\bar{j}}\ dz^{i}d\bar{z}^{j}
\eeq
and the complex structure two-form
\beq\label{b99}
\om = ig_{i\,\bar{j}}dz^{i} \wedge d\bar{z}^{j}
\eeq

The covariant constancy of J then implies :
\beq\label{b10}
(\ref{b5})\ \Rightarrow\ \ \Gamma^{i}_{\bar{j}k}\ =\
\Gamma^{i}_{\bar{j}\,\bar{k}}\ =\ \Gamma^{\bar{i}}_{jk}\ =\
\Gamma^{\bar{i}}_{j\bar{k}}\ =\ 0
\eeq
Using the well known expression of the symmetric connection
$\gamma^{\rho}_{\mu\nu}$ corresponding to the hermitian metric $g_{\mu\nu}$
\footnote{\ As usual, a comma indicates a derivative with respect to the
coordinate.} :
\beqa\label{b11}
\gamma^{i}_{jk} & = & {1\over2}g^{i\bar{l}}\ [g_{\bar{l}k,j}\ +\
g_{\bar{l}j,k}]\nnb \\
\gamma^{i}_{\bar{j}k}\ =\ \gamma^{i}_{k\bar{j}} & = & {1\over2}g^{i\bar{l}}\
[g_{\bar{l}k,\bar{j}}\ -\ g_{\bar{j}k,\bar{l}}]\nnb \\
\gamma^{i}_{\bar{j}\,\bar{k}} & = & 0
\eeqa
and conjugate expressions for $\gamma^{\bar{i}}_{\mu\nu}$, equations
(\ref{b10}) give :
\beqa\label{b12}
T_{ijk}\ =& T_{\bar{i}\,\bar{j}\,\bar{k}} &=\ 0 \nnb\\
T_{ij\bar{k}}\ =& g_{i\bar{k},j}\ -\ g_{j\bar{k},i} & \nnb\\
\Ga^{i}_{jk}\ =& \ga^{i}_{jk}\ -\ {1\over2}T^{i}_{jk} &=\
g^{i\bar{l}}g_{k\bar{l},j}\nnb\\
\Ga^{i}_{j\bar{k}}\ =&  -\ T^{i}_{j\bar{k}}\ =\ T^{i}_{\bar{k}j} &=\
g^{i\bar{l}}[g_{j\bar{l},\bar{k}}\ -\ g_{j\bar{k},\bar{l}}] \ = \
2\gamma^{i}_{j\bar{k}}\
\eeqa
Of course, the algebraic constraint (\ref{b6}) is identically satisfied.

The closedness of the torsion, equ. (\ref{b2}), then writes :
\beq\label{b13}
[T_{ij\bar{k},\bar{l}}\ -T_{ij\bar{l},\bar{k}}]+[T_{\bar{k}\,\bar{l}i,j}\
-T_{\bar{k}\,\bar{l}j,i}]\ =\ 0
\eeq
or equivalently :
$$ g_{i[\bar{k},\,\bar{l}\,]j}\    =\    g_{j[\bar{k},\,\bar{l}\,]i}    $$
Equation(\ref{b13}) can also be written :
\beq\label{b14}
D_{[\bar{l}}T_{\bar{k}\,]ij}\ +\ D_{[j}T_{i]\,\bar{k}\,\bar{l}}\ =\
T^{\nu}_{ij}T_{\nu\bar{k}\,\bar{l}} \ -\ 2T^{\nu}_{i[\bar{k}}T_{\bar{l}\,]\nu
j}\ \ \ \ \rm where\  \nu \equiv (n,\bar{n})\ \ .
\eeq
The closedness of the torsion may also be solved through the introduction of a
skew-symmetric torsion potential $b_{\mu\nu}$ (\cite{[33]}) :
$$ T_{\mu\nu\rho}\ =\ b_{\nu\rho ,\mu}\ +\ b_{\rho\mu ,\nu}\ +\ b_{\mu\nu
,\rho}\ $$
or, in complex coordinates
$$ T_{ij\bar{k}}\ =\ b_{j\bar{k} ,i}\ -\ b_{i\bar{k},j}\ +\ b_{ij ,\bar{k}}\ $$
Equation (\ref{b12}) then implies the existence of a vector potential
$(\chi_{i},\chi_{\bar{i}})$ and of a ``gauge" freedom $(v_{i},v_{\bar{i}})$
such that \cite{[33]},\cite{[3]} :
$$ b_{ij}\ =\ v_{j,i}\ -\ v_{i,j} $$
$$ b_{i\bar{j}}\ +\ g_{i\bar{j}}\ =\ \chi_{\bar{j},i}\ -\ v_{i,\bar{j}}$$
$$ $$

Due to equation (\ref{b10}), the Riemann tensor defined by
\beq\label{b15}
R^{\,\mu}_{\nu\rho\si}\ =\ \partial_{\rho} \Ga^{\mu}_{\nu\si}\ +\
\Ga^{\mu}_{\la\rho}\Ga^{\la}_{\nu\si}\ \ - \ (\rho \leftrightarrow \si)
\eeq
has many vanishing components in holomorphic coordinates
\beq\label{b16}
R^{\,i}_{\bar{j}\kappa\lambda}\ =\ R^{\,\bar{i}}_{j\kappa\lambda}\ =\ 0\ \ .
\eeq
The surviving ones are
\beqa\label{b17}
R^{i}_{jkl} & = & -g^{i\bar{n}}[\partial_{j} T_{kl\bar{n}}\ -\
2T_{\bar{n}[k}^{\bar{m}}g_{l\;]\bar{m},j}] \nnb\\
R^{i}_{jk\bar{l}}=-R^{i}_{j\bar{l}k} & = & -\partial_{\bar{l}} \Ga^{i}_{jk}\
-D_{k}T^{i}_{j\bar{l}} \ + \ T^{\bar{n}}_{\bar{l}k}T^{i}_{j \bar{n}} \nnb\\
R^{i}_{j\bar{k}\bar{l}} & = & g^{i\bar{n}}[\partial_{\bar{n}}
T_{\bar{k}\bar{l}j} -\ 2T_{j[\bar{k}}^{m}g_{\bar{l}\;]m,\bar{n}}]
\eeqa
and similar expressions for $R^{\,\bar{i}}_{\bar{j}\kappa\lambda}$.

The Ricci tensor
\beq\label{b18}
Ric_{\mu\nu}\ \stackrel{\rm def}{=}\ R^{\,\sigma}_{\mu\nu\sigma}
\eeq
then writes
\beqa\label{b19}
Ric_{ij} & = & D_{i}T^{k}_{kj}\ +\ T^{k}_{ij}T^{l}_{lk}\nnb\\
Ric_{i\,\bar{j}} & = & \partial_{i}\partial_{\bar{j}}\log \det\| g\| \ +\
D_{k}T^{k}_{i\bar{j}}\ -\ T^{k}_{i\bar{l}}T^{\bar{l}}_{\bar{j}k}
\eeqa
(for vanishing torsion, one recovers the usual results \cite{[4]}) .

We introduce the vector :
\beq\label{b20}
V_{\mu} = {1\over 2}J_{\mu}^{\nu}J_{\rho}^{\la}T^{\rho}_{\nu\la} \ \
\Leftrightarrow \left( \begin{array}{l}
V_{i}  =  T^{k}_{ki}\ =\ -T^{\bar{k}}_{\bar{k}i} \\
V_{\bar{i}}  =  T^{\bar{k}}_{\bar{k}\bar{i}}\ =\ -T^{k}_{k\bar{i}} \end{array}
\right)
\eeq
and obtain, by contraction of the closedness relation (\ref{b14}),
\beq\label{b201}
(D_{i}V_{\bar{j}}\ +\ D_{\bar{j}}V_{i})\ +\  (D_{k}T^{k}_{i\bar{j}}\ +\
D_{\bar{k}}T^{\bar{k}}_{\bar{j}i})\ -\ 2(T^{k}_{i\bar{j}}V_{k}\ +\
T^{\bar{k}}_{\bar{j}i}V_{\bar{k}}\ +\ T^{k}_{i\bar{l}}T^{\bar{l}}_{\bar{j}k})\
=\ 0
\eeq
and
\beq\label{b21}
g^{i\bar{j}}(D_{i}V_{\bar{j}}\ +\ D_{\bar{j}}V_{i})\ =\
g^{i\bar{j}}(T^{k}_{\bar{l}i}T^{\bar{l}}_{k\bar{j}}\ +\ 2V_{i}V_{\bar{j}})
\eeq
Moreover, as a consequence of equation (\ref{b12}), we find :
$$ D_{[\,\bar{l}}T_{\,\bar{k}\,]ij}\ =\  D_{[j}T_{i\,]\bar{k}\,\bar{l}} $$
which gives by contraction
\beq\label{b22}
D_{k}T^{k}_{i\bar{j}}\ -\  D_{\bar{k}}T^{\bar{k}}_{\bar{j}i}\ +\
D_{i}V_{\bar{j}}\ -\ D_{\bar{j}}V_{i}\ =\ 0\ \ .
\eeq
Using these equations, the Ricci tensor may be expressed in holomorphic
coordinates as :
\beqa\label{b24}
Ric_{ij} & = & D_{j}V_{i} + 2\partial_{[i}V_{j]} \nnb\\
Ric_{i\bar{j}}  & = &  D_{\bar{j}}V_{i} + {1\over
2}\left[\partial_{i}(\partial_{\bar{j}}\log \det\| g\| \ -\ 2V_{\bar{j}})\ +\
\partial_{\bar{j}}(\partial_{i}\log \det\| g\| \ -\ 2V_{i})\right] \ .
\eeqa

Finally, the scalar curvature
$$ R\ \stackrel{\rm def}{=}\ g^{\mu\nu} Ric_{\mu\nu} $$
writes :
\beq\label{b26}
R\ =\ 2g^{i\bar{j}}\left[ \partial_{i}\partial_{\bar{j}}\log \det\| g\| \ +\
2V_{i}V_{\bar{j}}\ -\ {1\over 2}(D_{i}V_{\bar{j}}\ +\ D_{\bar{j}}V_{i})\right]
\eeq

\subsection{Generalised Quasi Ricci flat metrics}
For further use, we recall that in \cite{[8]}, Friedan defines ``quasi Ricci
flat metrics " through (no torsion)
$$ \exists\ \ V_{\mu} \ \ \ \   {\rm such\ that} \ \ \ \    Ric_{\mu\nu} =
D_{\mu}V_{\nu}\ +\  D_{\nu}V_{\mu} $$
Such geometries lead to one-loop finiteness for the D=2 non-linear $\si$ models
built on such target space metrics. They have been studied by Bonneau and
Delduc and some explicit expressions have been obtained  \cite{[55]}.
Here, in the presence of torsion, the Ricci tensor is no longer symmetric and,
following Friedling and van de Ven \cite{[9]}, we define ``generalised quasi
Ricci flatness " by the same one loop finiteness requirement. This leads to the
definition :

\beqa\label{b27}
\exists\ \ W_{\mu}\ {\rm and \ }\ \chi_{\mu} \ \   {\rm such\ that} \ \
Ric_{(\mu\nu)} & = &  \nabla_{(\mu}W_{\nu)} \nnb \\
 Ric_{[\mu\nu]} & = & {1\over 2}T^{\rho}_{\mu\nu}W_{\rho} +
\nabla_{[\mu}\chi_{\nu]}
\eeqa
In this (2,0) heterotic geometry, conditions (\ref{b27}) simplify to :
\beqa\label{b28}
Ric_{ij} & = & D_{j}W_{i} + \partial_{[i}(W+\chi)_{j]} \nnb\\
Ric_{i\bar{j}}  & = &  D_{\bar{j}}W_{i} + \partial_{[i}(W+\chi)_{\bar{j}]}\ .
\eeqa

\subsection{SU(N) holonomy and generalised quasi Ricci flatness}
Let us recall that the holonomy group of 2N dimensional (with or without
torsion) manifolds with a covariantly constant complex structure is a subgroup
of U(N). Moreover, particular cases where the holonomy is SU(N) play a special
role : in the absence of torsion, this means Ricci flatness ; in the present
case, the vanishing of the U(1) part of the Riemann curvature $$C_{\mu\nu} =
J_{\rho}^{\si}R^{\rho}_{\si\mu\nu}$$ writes :
\beq\label{b31}
C_{\mu\nu} = 2\partial_{\;[\mu}\Ga_{\nu]} = 0
\eeq
where $\Ga_{\mu} = J_{\rho}^{\nu}\Ga^{\rho}_{\nu\mu}$, ({\sl a priori } not a
vector), is found in holomorphic coordinates as \footnote{\ We use equations
(\ref{b17},\ref{b12},\ref{b20})\ .}\cite{[35]} :
\beq\label{b32}
\Ga_i = \ i[\partial_i\log \det\|g\| - 2V_i] \ \ ,\  \Ga_{\bar{i}} =
-i[\partial_{\bar{i}}\log \det\|g\| - 2V_{\bar{i}}]\ \ .
\eeq
Using the tensor $C_{\mu\nu}$, equation (\ref{b24}) writes
\beq\label{b33}
Ric_{ij} = D_jV_i + {i\over 2}\,C_{ij}\ \ ,\ \ Ric_{i\bar{j}} = D_{\bar{j}}V_i
+ {i\over 2}\,C_{i\bar{j}}
\eeq
Then SU(N) holonomy $\Leftrightarrow C_{\mu\nu} = 0\ $, leads to a special case
of ``generalised quasi Ricci flatness" (\ref{b28}) with $$W_i = V_i \ \ ,\
\chi_{i} = - V_i \ .$$ As a consequence, and as will be exemplified in
subsection (3.3.2), SU(N) holonomy is a more restrictive requirement that the
sole ``generalised quasi Ricci flatness" of Friedling and van de Ven.

\subsection{(2,0) versus (2,2) geometries}
Given a complex Riemannian manifold  - $J_{\mu}^{\nu}$ being the complex
structure - equipped with a metric $g_{\mu\nu} $ and a torsion tensor
$T_{\mu\nu\rho}$ satisfying equations
(\ref{b1},\ref{b2},\ref{b3},\ref{b4},\ref{b7},\ref{b5}), we address the
following question :

{\sl ``does there exist a conformal transformation that transforms this (2,0)
geometry into a (2,2) one "}, i.e. that relates the metric $g_{\mu\nu} $ and
the torsion tensor $T_{\mu\nu\rho}$ to a K\"{a}hler metric $\hat{g}_{\mu\nu}$
(no torsion) through :
\beq\label{b261}
\hat{g}_{\mu\nu} = e^{-2f}g_{\mu\nu}
\eeq
(the positivity of the distance requires the reality of f ).

\noindent The K\"{a}hler condition writes in complex coordinates :
$$\hat{g}_{i\bar{j} ,k} = \hat{g}_{k\bar{j},i} \ \ , \ \ \ \hat{g}_{i\bar{j}
,\bar{k}} = \hat{g}_{i\bar{k},\bar{j}} $$
which leads, using equation (\ref{b12}), to
\beqa\label{b262}
T_{ij\bar{k}}\ & = & g_{i\bar{k},j}\ -\ g_{j\bar{k},i} = 2(
g_{i\bar{k}}\partial_j f\ -\ g_{j\bar{k}}\partial_i f) \nnb \\
T_{\bar{i}\,\bar{j}k}\ & = & g_{\bar{i}k,\bar{j}}\ -\ g_{\bar{j}k,\bar{i}} = 2(
g_{\bar{i}k}\partial_{\bar{j}} f\ -\ g_{\bar{j}k}\partial_{\bar{i}} f)
\eeqa
The vector $V_{\mu}$ is then a gradient :
\beq\label{b264}
V_{i} = T^{k}_{ki} = 2(N-1)\partial_{i} f \ \ ,
V_{\bar{i}} = T^{\bar{k}}_{\bar{k}\bar{i}} = 2(N-1)\partial_{\bar{i}} f \ .
\eeq
This is a very restrictive condition as it means that the whole torsion tensor
depends on a single real function f. Notice that equation (\ref{b262}) may be
rewritten as :
\beq\label{b2622}
T_{ij\bar{k}}\ = {1\over (N-1)}\left[ g_{i\bar{k}}V_j - g_{j\bar{k}}V_i
\right]\ \ \ ,\ \ T_{\bar{i}\,\bar{j}k}\ = {1\over (N-1)}\left[
g_{\bar{i}k}V_{\bar{j}} - g_{\bar{j}k}V_{\bar{i}} \right]\
\eeq
The closedness relation (\ref{b21}) writes
\beq\label{b263}
D_{i}V^{i}\ +\ D_{\bar{j}}V^{\bar{j}}\ =\ 2\frac{N-2}{N-1}V_iV^i  \ \ {\rm
where\ \ } V^i = g^{i \bar{j}} V_{\bar{j}} \ \ ,  V^{\bar{j}} = g^{i \bar{j}}
V_{i} \ \ .
\eeq
Equation (\ref{b22}) gives $D_{i}V^{i}\ =\ D_{\bar{i}}V^{\bar{i}}\  $
and, for $N \neq 2$,  $D_{i}V_{\bar{j}}\ =\ D_{\bar{j}}V_{i}\  \ .$
As a consequence, the conformal factor $e^{2f}$ satisfies :
\beqa\label{b265}
\hat{\Delta}e^{2f} & \stackrel{\rm def}{=} &
2\hat{g}^{i\bar{j}}\hat{\nabla}_i\partial_{\bar{j}}e^{2f} =
2\hat{g}^{i\bar{j}}\partial_i\partial_{\bar{j}}e^{2f} \ = \  4
\hat{g}^{i\bar{j}}e^{2f}\left[\partial_i\partial_{\bar{j}}f +
2\partial_{i}f\partial_{\bar{j}}f \right] \nnb\\
& = & 4e^{4f}g^{i\bar{j}}\left[D_i\partial_{\bar{j}}f - V_i\partial_{\bar{j}}f
+ {1\over{(N-1)}}V_i\partial_{\bar{j}}f\right] \nnb\\
& = & \frac{2e^{4f}}{N-1}\left[D_i V^i - \frac{N-2}{N-1}V_i V^i\right] = 0\ .
\eeqa
and we have also
\beq\label{b2651}
\Delta f \stackrel{\rm def}{=} 2g^{i\bar{j}}D_i\partial_{\bar{j}}f =
\frac{8(N-2)}{(N-1)^2}\|V\|^2.
\eeq
Notice that using (\ref{b99}, \ref{b262}) one has
\beq\label{bb2}
d\omega=2\omega\wedge df.
\eeq

Finally, the scalar curvatures are related by :
\beq\label{b266}
R[g,T]\ =\ e^{-2f}\left[\hat{R}[\hat{g},\hat{T}\equiv 0] +
\frac{2N(N-2)}{(N-1)^2}\|V\|^2_{\hat{g}}\right]
\eeq

\section{(2,0) heterotic geometry in 2 complex dimensions}
\subsection{The torsion tensor}
In that special case $T_{ij\bar{k}} $ has as many components as the vector
$V_{i}$ and from definition (\ref{b20}) one obtains :
\beq\label{c1}
T_{ij\bar{k}}\ =\ g_{i\bar{k}}V_{j}\ -\  g_{j\bar{k}}V_{i}
\eeq
This relation is a duality one, first written in real coordinates in \cite{[6]}
$$T_{\mu\nu\rho}\ =\ \epsilon _{\mu\nu\rho\sigma}V^{\sigma}$$
The closedness relation (\ref{b21}), when compared to equation (\ref{b22})
gives
\beq\label{c2}
D_{i}V^{i}\ =\ D_{\bar{j}}V^{\bar{j}}\ =\ 0
\eeq
and the scalar curvature reduces to :
\beq\label{c3}
R\ =\ 2g^{i\bar{j}}( \partial_{i}\partial_{\bar{j}}\log \det\| g\| \ +\
2V_{i}V_{\bar{j}})
\eeq

\subsection{ From (2,2) to (2,0) supersymmetry through a conformal rescaling of
the metric ?}

When one compares equations (\ref{c1}) and (\ref{b264},\ref{b2622}), one sees
that in 2 complex dimensions one gets conformal equivalence between (2,0) and
(2,2) geometries, if and only if the vector V is the gradient of a real
function :
\beq\label{c33}
V_j = 2\partial_j f \ \ ,\ V_{\bar{j}} = 2\partial_{\bar{j}}f
\eeq
whith the constraints
\beq\label{c333}
\hat{\De}e^{2f} = \De f = 0
\eeq
and, in that case, the scalar curvatures are proportional :
\beq\label{c34}
R[g,T]\ =\ e^{-2f}\hat{R}[\hat{g},\hat{T}\equiv 0]
\eeq
$$ $$

We now construct somes examples of (2,0) heterotic metrics with particular
isometries.
\subsection{Special cases with linear U(N) isometry}
\subsubsection{Generalities}
Let us consider the special family of hermitian metrics $g_{i\bar{j}}\
(i,\bar{j} = 1,..,N)$ with linear U(N) symmetry
\beq\label{c4}
g_{i\bar{j}}\ =\ A(s)\delta_{ij}\ +\ B(s)\bar{z}^{i}z^{j}\ \ \quad
\ , \ s = \sum_{i=1}^{N}\bar{z}^{i}z^{i}
\eeq
Defining $C(s) = A(s) + sB(s)$ we have
\beq\label{c5}
g^{i\bar{j}}\ =\ {1\over A(s)}\delta_{ij}\ -\
\left[\frac{B(s)}{A(s)C(s)}\right]z^{i}\bar{z}^{j}
\eeq
and $ \det\|g\| = A(s)C(s)$. From (\ref{b12}), the torsion tensor is found to
be \footnote{\  A dot indicates a derivative with respect to the variable s. }
:
\beq\label{c6}
T_{ij\bar{k}}\ =\ [\dot{A}(s)\ -B(s)](\bar{z}^{j}\delta_{ik}\ -\
\bar{z}^{i}\delta_{jk})
\eeq
and its closedness gives
\beq\label{c7}
\frac{d}{ds}{[\dot{A}(s)\ -B(s)]}\delta_{i}^{[\,j}z^{k\,]}\bar{z}^{l}\ +\
[\dot{A}(s)\ -B(s)]\delta_{i}^{[\,j}\delta_{l}^{k\,]}\ \ \ -\ \ (i
\leftrightarrow l)\  =\ 0
\eeq
For  N$ \geq{3}$, this implies $\dot{A}(s)\ -B(s)\ =\ 0$ which means that the
metric is K\"ahler and that no torsion can be put on a manifold with such
isometries. For N = 1, the torsion tensor identically vanishes whereas for
N = 2 (2 complex dimensions), equation (\ref{c7}) gives
\beq\label{c8}
\frac{d}{ds}[s^{2}(\dot{A}(s)\ -B(s))]\ =\ 0\ \ \Rightarrow \ \
\dot{A}(s)\ -B(s)\ =\ {L\over{2s^{2}}}
\eeq
In such a case, the vector $V_{i}$ writes
\beq\label{c9}
V_{i}\ =\ T^{k}_{ki}\ =\ \left[\frac{\dot{A}(s)\
-B(s)}{A(s)}\right]\bar{z}^{i}\  \stackrel{\rm def}{=} \partial_{i}\log \gamma
(L,s)
\eeq
Due to the reality of the functions A(s) and B(s) and consequently of
$\ga(L,s)$, the looked-for metric is conformally equivalent to a K\"{a}hler one
through the conformal transformation  $\hat{g} = \ga(L,s)g$ (compare equations
(\ref{c33}) and (\ref{c9})).
We then rescale the looked-for functions $A(s)$ and $C(s)$ according to :
\beq\label{c91}
A(s) = \ga (L,s)\eta (s) \ \ ,\ C(s) = \ga (L,s)\mu (s)
\eeq
where, as a consequence of equations (\ref{c8},\ref{c9}) we have :
\beq\label{c92}
\mu(s) \ = \ \frac{d(s\eta(s))}{ds} \ \ ,\ \ s\frac{d\ga}{ds} \ =  \
\frac{L}{2s\eta} \ \ .
\eeq

In this U(2) case, it is convenient to use the coordinates
$$ z^1=\sqrt{s}\cos\frac{\theta}{2} e^{i\frac{(\phi+\psi)}{2}}\ \ ,\ \
 z^2=\sqrt{s}\sin\frac{\theta}{2}e^{i\frac{(\phi -\psi)}{2}}$$
to write the distance (\ref{b9}) :
\beq\label{c921}
d\tau^2 = \frac{C(s)}{2s}(ds)^2 + 2sA(s)(\eta_1^2+\eta_2^2) + 2sC(s)\eta_3^2
\eeq
where the expressions for the $\eta_i$ are given in subsection (5.1).

\noindent The conformal rescaling then gives :
\beq\label{c922}
d\tau^2 = \ga(L,s)\left\{\frac{1}{2s}\frac{d\si(s)}{ds}(ds)^2 +
2\si(s)(\eta_1^2+\eta_2^2) + 2s\frac{d\si(s)}{ds}\eta_3^2\right\}
\eeq
where $\si(s) = s\eta(s)$.

As mentionned in the Introduction, despite the large isometry group considered,
(2,0) supersymmetry is not sufficient to fix the geometry of the manifold :
other constraints are needed. Three of them will be considered in this work :

- more supersymmetries : see next section,

-`` generalised quasi Ricci flatness " as defined in the previous section,

- scalar-curvature flatness.

\subsubsection{Generalised Quasi Ricci flat metrics}

Due to the linear U(N) isometry, the Ricci tensor is symmetric and, when
compared to equations (\ref{b24}), conditions (\ref{b28}) write :
\beq\label{c10}
D_j[W_i - V_i] = \ 0
\eeq
and :
\beq\label{c101}
D_{\bar{j}}[W_i - V_i] = \partial_i \partial_{\bar{j}}\log
\frac{\det\|g\|}{\ga^2(L,s)}
\eeq
Using equations (\ref{c8},\ref{c9}), (\ref{c10}) gives :
$$ W_i = V_i + \kappa C(s)\bar{z}^i$$
Equation (\ref{c101}) then leads to :
\beq\label{c1011}
\kappa\frac{d(sA(s))}{ds} = \frac{d}{ds}\log(A(s)C(s)) - \frac{L}{s^2A(s)}
\eeq

Under the conformal rescaling (\ref{c91}), this differential equation gives,
after a first integration  \footnote{\ An inessential integration constant has
been suitably chosen.}:
\beq\label{c102}
\frac{d\si(s)}{ds} = \frac{s}{\si(s)} \exp (\kappa\ga(L,s)\si(s)) \ \ .
\eeq
We were not able to solve explicitly the system (\ref{c92},\ref{c102}), but the
distance (\ref{c922}) writes :
\beq\label{c103}
d\tau^2 = \ga(L,s)\left\{\frac{\si}{2s^2(\si)}\exp (-\kappa\ga\si)(d\si)^2 +
2\si(\eta_1^2+\eta_2^2) + \frac{2s^2(\si)}{\si}\exp
(\kappa\ga\si)\eta_3^2\right\}
\eeq
and the geometrical objects are :
\beqa\label{c105}
V_{i}\ =\ \partial_{i}\log \ga (L,s) &   , &  \det\| g\| \ =\ \ga^{2}(L,s)\exp
(\kappa\ga(L,s)\si(s))/4 \nnb\\
Ric_{ij}  =  D_{i}V_{j}\ =\ D_{j}V_{i} & , &
Ric_{i\bar{j}}  =  D_{\bar{j}}V_{i}\ +\ {\kappa^2\over 2}
\partial_{i}\partial_{\bar{j}}[\ga(L,s)\si(s)] .
\eeqa

In the special case $\ka \ =\ 0$, equations (\ref{c10},\ref{c101}) give with
(\ref{b32},\ref{c9}):
$$W_i = V_i\ \ ,\ \Ga_i = \Ga_{\bar{i}} = 0$$ and one gets SU(2) $\equiv $
Sp(1) holonomy. The isometries then enforce the solution to be Eguchi-Hanson
metric with torsion \footnote{\  which is discussed in subsect. (5.1)\ .} :

\beqa\label{c11}
A(s)  =  \frac{\ga (L,s)}{2s}\sqrt{s^{2}+\la^{2}} & , &
C(s) =  \frac{\ga (L,s)}{2s}\frac{s^{2}}{\sqrt{s^{2}+\la^{2}}} \nnb\\
s\frac{d}{ds}\ga(L,s) & = & \frac{L}{\sqrt{s^{2}+\la^{2}}}
\eeqa
L and $\la^{2}$ being constants and
\beq\label{c122}
V_{i}\ =\ \partial_{i}\log \ga (L,s) \   ,\ \ \ \det\| g\| \ =\
\frac{\ga^{2}(L,s)}{4}\ ,\ \ \ R\ =\ 0\ .
\eeq
This illustrates how one-loop finiteness, i. e. generalised quasi Ricci
flatness is a less restrictive requirement than SU(N) holonomy.
$$ $$

For vanishing torsion ( L=0 ),\  $\ga$ is a constant which may be taken to be
1, and we recover the metric first derived by Bonneau and Delduc \cite{[55]}.
Equation (\ref{c102}) integrates to :
\beq\label{c123}
s^2(\si)=\displaystyle\frac 2{\ka^2}\left[1-(1+\ka\si)e^{-\ka\si}\right]-d
\eeq
where d is a real integration constant. This metric has a non-vanishing scalar
curvature
$$ R=4\ka\left(1+\frac{\ka s^2(\si) e^{\ka\si}}{2\si}\right).$$
The distance (\ref{c103}) specifies to :
$$\displaystyle d\tau^2 = \frac{\si e^{-\ka\si}}{2s^2(\si)}(d\si)^2+
2\si(\eta_1^2+\eta_2^2)+\frac{2s^2(\si)}{\si e^{-\ka\si}}\eta_3^2 \ \ . $$
Using this form of the distance, one can check that it is indeed regular for
$\ka\leq 0$ and $d>0.$ We take for variable $\si\in[\si_0,+\infty[$ where
$\si_0$ is defined by $s(\si_0)=0.$ For $ \si\to +\infty$ this distance is
asymptotically flat ( similarly to the torsionless Taub-NUT ) and it exhibits a
bolt n=2 for $\si\to\si_0$ ( similarly to the torsionless Eguchi-Hanson ). This
means that, starting for $\ka =0$ from an asymptotically locally euclidean
metric we get, for $\ka<0$, an asymptotically locally flat metric.

\subsubsection{Metrics with a vanishing scalar curvature}
When the isometry group is U(2), equation (\ref{c3}) gives
\beq\label{c13}
R\ =\ 2\left[\frac{D + s\dot{(D)}}{C}\ +\ \frac{D}{A}\ +\
\frac{L^{2}}{2s^{3}A^{2}C}\right] \ \ {\rm where }\ \ D \stackrel{\rm def}{=}
\frac{d}{ds}\log \det\| g\|\ \ .
\eeq

After the conformal rescaling (\ref{c91}), the condition R = 0 becomes a L
independent third order differential equation  :
$$ s\frac{d}{ds}\left(\frac{\ddot{\si}}{\dot{\si}}\right) +
\frac{\ddot{\si}}{\dot{\si}} + 2s\frac{\ddot{\si}}{\si} = 0 \ \ $$
which integrates in a first step to
$$\si^2\left(s\frac{\ddot{\si}}{\dot{\si}}\right) = c $$
and then the looked-for metric is given by :
\beqa\label{c14}
A(s) = \frac{\ga (L,s)}{s}\si (s) \ & , & \ C(s) =  \frac{\ga
(L,s)}{s}\frac{[\si ^{2}(s) + 2c\si (s)+ d]}{\si(s)}  \nnb\\
s\frac{d}{ds}\ga (L,s) = \frac{L}{2\si (s)} \ & , & \ s\frac{d}{ds}\si (s)  =
\frac{[\si ^{2}(s) + 2c\si(s) + d]}{ \si (s)}
\eeqa
L, c and d being constants. The geometrical objects are :
\beqa\label{c15}
V_{i}\ =\ \partial_{i}\log \ga (L,s) &   , &  \det\| g\| \ =\
\frac{\ga^{2}(L,s)[\si ^2 + 2c\si + d]}{s^2} \nnb\\
Ric_{ij} = D_{i}V_{j}\ =\ D_{j}V_{i} \ & , & \ Ric_{i\bar{j}} = Ric_{\bar{j}i}
= D_{\bar{j}}V_{i}\ +\ \partial_{i}\partial_{\bar{j}}\log \frac{[\si ^2 + 2c\si
+ d]}{s^2} \ \ .
\eeqa

As announced in subsection (3.3.1), these metrics are obtained through a
conformal rescaling $\ga (L,s)$ of the K\"{a}hler, scalar flat \footnote{\
scalar flatness being conserved through a conformal transformation in 2 complex
dimensions (\ref{c34}).}, torsionless ones which are known as LeBrun metrics
\cite{[44]}. These last ones are asymptotically euclidean and regular
for suitable choices of the parameters $c$ and $d$. They give counterexamples
to the Generalised Positive Action conjecture of Hawking and Pope \cite{[45]}.
On the contrary, for L $\neq$ 0, our metrics (\ref{c14}), although still
asymptotically euclidean, are always singular : we have checked this on the
expression of the curvature tensor.

$$  \ $$
In the special case c = 0, one recovers Eguchi-Hanson metric with
torsion ($d = -\la^{2}$)
\beqa\label{c144}
A(s) = \frac{\ga (L,s)}{2s}\sqrt{s^{2}+\la^{2}} \ & , & \ C(s) = \frac{\ga
(L,s)}{2s}\frac{s^2}{\sqrt{s^{2}+\la^{2}}} \nnb\\
s\frac{d}{ds}\ga (L,s) & = & \frac{L}{\sqrt{s^{2}+\la^{2}}}\ \ .
\eeqa

Let us emphasize that we have obtained a new and easy derivation of
Eguchi-Hanson metric with torsion directly in complex coordinates.

\subsection{More examples : Calabi metrics with torsion}
\subsubsection{ Generalities}
In \cite{[10]} Calabi exhibits K\"ahler torsionless metrics in 2N real
dimensions.
He considers the special family of metrics with a linear O(N) symmetry and
depending only on N coordinates :
\beqa\label{c16}
x_{i} &  = & z^{i}\ +\ \bar{z}^{i} \ \ \ \ i=1,..N \ \ \quad
\ ,\ \ s = \sum_{i=1}^{N} x_{i}^{2}\nnb\\
g_{i\bar{j}} & = & A(s)\delta_{ij}\ +\ B(s)x_{i}x_{j}
\eeqa
Then, with $C(s) = A(s) + sB(s)$ we have
\beq\label{c55}
g^{i\bar{j}}\ =\ {1\over A(s)}\delta_{ij}\ -\
\left[\frac{B(s)}{A(s)C(s)}\right]x_{i}x_{j}
\eeq
and $\det\|g\| = A(s)C(s)$. From (\ref{b12}), the torsion tensor is found to be
\footnote{\ A dot indicates a derivative with respect to the variable s.}
\beq\label{c17}
T_{ij\bar{k}}\ =\ [2 \dot{A}(s)\ -B(s)](x_{j}\delta_{ik}\ -\ x_{i}\delta_{jk})
\eeq
and its closedness gives
\beq\label{c18}
2\frac{d}{ds}[2\dot{A}(s)\ -B(s)]\delta_{i[\,j}x_{k\,]}x_{l}\ +\ [2\dot{A}(s)\
-B(s)]\delta_{i[\,j}\delta_{k\,]\,l}\ \ \ -\ \ (i \leftrightarrow l)\  =\ 0
\eeq
For  N$ \neq{2}$, this implies a vanishing torsion tensor. On the contrary, for
N = 2 (2 complex dimensions), equation (\ref{c18}) gives :
\beq\label{c19}
\frac{d}{ds}\left[s(2\dot{A(s)}\ -B(s))\right]\ =\ 0 \ \ \Rightarrow \ \
2\dot{A(s)}\ -B(s)\ =\ {L\over{s}}
\eeq
In such a case, the vector $V_{i}$ writes
\beq\label{c20}
V_{i}\ =\ T^{k}_{ki}\ =\ \left[\frac{2\dot{A}(s)\ -B(s)}{A(s)}\right]x_{i}\
\stackrel{\rm def}{=} \partial_{i}\log \ga (L,s)
\eeq

Here again, due to the reality of the functions A(s) and B(s) and consequently
of $\ga(L,s)$, the looked-for metric is conformally equivalent to a K\"{a}hler
one through the conformal transformation  $\hat{g} = \ga(L,s)g$ (compare
equations (\ref{c33}) and (\ref{c20})).
We then rescale the looked-for functions $A(s)$ and $C(s)$ according to :
\beq\label{c911}
A(s) = \ga (L,s)\eta (s) \ \ ,\ C(s) = \ga (L,s)\mu (s)
\eeq
where, as a consequence of equations (\ref{c19},\ref{c20}) :
\beq\label{c912}
\mu(s)\ \ = \eta(s)s\frac{d}{ds}\log [s\eta^2(s)] \ \ ,\ \ s\frac{d\ga}{ds} \ =
 \ \frac{L}{2\eta} \ \  .
\eeq
We now add further geometrical constraints.

\subsubsection{Generalised Quasi Ricci flat metrics}

Due to the linear O(N) isometry, the Ricci tensor is symmetric and then
equations (\ref{c10},\ref{c101}) hold.
Using equations (\ref{c19},\ref{c20}), we obtain :
$$ W_i = V_i $$
and :
\beq\label{c21}
\log \frac{\det\| g\|}{\ga^{2}(L,s)}\ =\ {\rm constant}\ \ \Leftrightarrow \
\frac{d}{ds}(s\eta^2)\ \ =\ {\rm constant}\
\eeq
Then, the looked-for metric here also has SU(2) $\equiv$ Sp(1) holonomy and
writes :
\beqa\label{c22}
A(s) = \frac{\ga (L,s)}{2\sqrt{s}}\sqrt{s+\la} \ & , &\
C(s) = \frac{\ga (L,s)}{2\sqrt{s}}\frac{s}{\sqrt{s+\la}} \nnb\\
s\frac{d}{ds}\ga (L,s) & = & L\sqrt{\frac{s}{s+\la}}
\eeqa
L and $\la$ being constants and
\beqa\label{c23}
V_{i}\ =\ \partial_{i}\log \ga (L,s) \  & ,& \ \ \det\| g\| \ =\
\frac{\ga^{2}(L,s)}{4}\nnb\\
Ric_{ij} = \ D_i V_j = \ D_{j}V_{i} \ &, & \
Ric_{i\bar{j}} = Ric_{\bar{j}i} = D_{\bar{j}}V_{i}\ =\ D_{i}V_{\bar{j}} \ \ ,\
\ R\ =\ 0\ .
\eeqa

\subsubsection{Metrics with a vanishing scalar curvature}
In this case, equation (\ref{c3}) gives
\beq\label{c131}
R\ =\ 4\left[\frac{D + 2s\dot{D}}{C}\ +\ \frac{D}{A}\ +\
\frac{L^{2}}{sA^{2}C}\right] \ \ {\rm where }\ \ D \stackrel{\rm def}{=}
\frac{d}{ds}\log \det\| g\|\ \ \ .
\eeq
After the conformal rescaling (\ref{c911}), the condition R = 0 becomes a L
independent differential equation \footnote{\ $\ddot{(s\eta^2)} =
\frac{d}{ds}\dot{(s\eta^2)}= \frac{d}{ds}\left[\frac{d}{ds}(s\eta^2)\right]\
$}:
$$ \frac{\ddot{(s\eta^2)}}{\dot{(s\eta^2)}}\left[ 1 + s\frac{d}{ds}\log \eta
\right] + s\frac{d}{ds}\left[\frac{\ddot{(s\eta^2)}}{\dot{(s\eta^2)}}\right] =
0$$
which integrates in a first step to
$$s\eta\frac{\ddot{(s\eta^2)}}{\dot{(s\eta^2)}} = \kappa$$
Taking $u = \dot{(s\eta^2(s))} $ as a new variable, one is led to a Riccati
equation :
$$ 2\kappa\frac{d\eta}{du} + {1\over u}\eta^2 = 1 \ \ . $$
We then find two families of solutions which hereagain are obtained from the
torsionless case (L=0) through a conformal rescaling of the metric :

- the quasi Ricci flat one (\ref{c22}) for $\kappa = 0$,
$$\ $$

- a new one, given through the following equations :
\beqa\label{c24}
A[v(s)] & = & v\left[ \ga_1 + L \log v \right]\frac{I_1(v) - d K_1(v)}{ I_0(v)
+ d K_0(v)} \nnb\\
C[v(s)] & = &  v\left[ \ga_1 + L \log v \right]\frac{ I_0(v) + d K_0(v)}{
I_1(v) - d K_1(v)} \eeqa
and where $s  =  c^2[I_0(v) + d K_0(v)] $ defines the function $v(s)$ ; L,
$\ga_1$, c and $d$ are constants and $I(v)$ and $K(v)$ the usual Bessel
functions. Moreover,
\beqa\label{c25}
V_{i}\ =\ \partial_{i}\log [ \ga_1 + L \log v] &   , &  \det\| g\| \ =\
v^2\left[ \ga_1 + L \log v \right]^2 \nnb\\
Ric_{ij} =  D_{i}V_{j}\ =\ D_{j}V_{i} & , & Ric_{i\bar{j}}  = Ric_{\bar{j}i} =
D_{\bar{j}}V_{i}\ +\ \partial_{i}\partial_{\bar{j}}\log v^2(s)\ \ .
\eeqa

$$ \ $$

\section{(4,0) heterotic geometry}
We add to conditions (\ref{b2}-\ref{b5}) for (2,0)  supersymmetry, the
following ones :

- apart from the first complex structure, labelled as $J_{3\mu}^{\nu}$, there
does exist another one, $J_{1\mu}^{\nu}$, integrable, covariantly constant and
anticommuting with $ J_{3\mu}^{\nu}$ :
\beq\label{D1}
 J_{1\mu}^{\nu} J_{1\nu}^{\rho} = - \delta_{\mu}^{\rho}
\eeq
\beq\label{D2}
N^{\rho}_{1\mu\nu}\ \equiv \  J_{1\mu}^{\la}(\partial_{\la} J_{1\nu}^{\rho}\ -\
\partial_{\nu} J_{1\la}^{\rho})\ -(\mu\ \leftrightarrow \ \nu)\ =\ 0
\eeq
\beq\label{D3}
D_{\mu}J_{1\nu}^{\rho} \ =\ 0
\eeq
\beq\label{D4}
J_{1\mu}^{\nu}J_{3\nu}^{\rho} \ =\ -J_{3\mu}^{\nu}J_{1\nu}^{\rho}
\eeq

- hermiticity : the metric is also hermitian with respect to the complex
struture $J_{1\mu}^{\nu}$
\beq\label{D5}
g_{\mu\la}J_{1\nu}^{\la} \ + \ J_{1\mu}^{\la}g_{\la\nu} \ =\ 0
\eeq
As a consequence (see for example ref.\cite{[77]}),
\beq\label{D6}
J_{2\mu}^{\nu} \stackrel{\rm def}{=}\  J_{3\mu}^{\rho} J_{1\rho}^{\nu}
\eeq
is a third complex structure, integrable and covariantly constant ; moreover
the metric is also hermitian with respect to $J_{2\mu}^{\nu}$ and the triplet
of complex strutures  $J_{a\mu}^{\nu}$ (a = 1,2,3) satisfies a quaternionic
multiplication law :
\beq\label{D7}
J_{a\mu}^{\nu} J_{b\nu}^{\rho}\ =\ -\delta_{\mu}^{\rho}\ +\ \epsilon_{abc}
J_{c\mu}^{\rho}\ \ \ {\rm with  }\ \ \epsilon_{123} \ =\ +1
\eeq
Moreover, the generalized Nijenhuis tensors
\beq\label{D8}
N^{\rho}_{ab\mu\nu}\ \equiv \  [J_{a\mu}^{\la}(\partial_{\la} J_{b\nu}^{\rho}\
-\ \partial_{\nu} J_{b\la}^{\rho})\ -(\mu\ \leftrightarrow \ \nu)] \ +\ (a
\leftrightarrow b)
\eeq
vanish, which ensures the N=4 supersymmetry algebra \cite{[77]}.

\noindent The sufficient character of equations (\ref{D1}-\ref{D5}) is often
missed in the literature where (\ref{D7},\ref{D8}) are added as independent
conditions although they are direct consequences of the others.

The dimension of the manifold has to be a multiple of 4, and we shall now
translate these new conditions in complex coordinates adapted to the complex
struture $J_{3}$ (see  (\ref{b8})). As a consequence of equation (\ref{D7}),
all components of the tensors :
\beq\label{D9}
K_{\mu}^{\nu} =  \frac{1}{2}(J_{1}-iJ_{2})_{\mu}^{\nu} \ \  , \ \
\bar{K}_{\mu}^{\nu}  =  \frac{1}{2}(J_{1}+iJ_{2})_{\mu}^{\nu}
\eeq
vanish, but for
\beq\label{D10}
K_{i}^{\bar{j}}  =  J_{1i}^{\bar{j}}  =  -iJ_{2i}^{\bar{j}} \ \  , \ \
\bar{K}_{\bar{i}}^{j}  =  J_{1\bar{i}}^{j}  = \  iJ_{2\bar{i}}^{j}
\eeq
with $ i,j,\bar{i},\bar{j} = $1,2,...2N. The hermiticity of the metric then
implies the skew-symmetry of $ K_{\mu\nu}$ and $\bar{K}_{\mu\nu}$ which
therefore locally define (2,0) and (0,2) forms :
\beqa\label{D11}
\om_1 - i\om_2 = & (K_{i}^{\bar{j}}g_{\bar{j}k})dz^{i}\wedge dz^{k} & =
J_{1ik}dz^{i}\wedge dz^{k}\nnb\\
\om_1 + i\om_2 = & (\bar{K}_{\bar{i}}^{j}g_{j\bar{k}})dz^{\bar{i}}\wedge
dz^{\bar{k}} & = J_{1\bar{i}\,\bar k}dz^{\bar{i}}\wedge dz^{\bar{k}}
\eeqa
In this coordinate system, the covariant constancy of the complex structures
$J_{1}$ and $J_{2}$ implies :
\beq\label{D12}
\partial_{\bar{i}}K_{jk}  =  T^{l}_{\bar{i}j}K_{lk}\ -\  T^{l}_{\bar{i}k}K_{lj}
\ \ ,\ \ \partial_{i}K_{jk} = \Ga^{l}_{ji}K_{lk}\ -\  \Ga^{l}_{ki}K_{lj}
\eeq
and the complex conjugate equations.
Multiplying equations (\ref{D12}) by $(K^{-1})^{jk} = (\bar{K})^{jk}$ gives :
\beqa\label{D121}
{1\over 2}(K^{-1})^{jk}\partial_{\bar{i}}K_{jk} & = & -V_{\bar{i}} \nnb\\
{1\over 2}(K^{-1})^{jk}\partial_{i}K_{jk} \ =\ -\Ga^{l}_{li} & = & V_{i} -
\partial_{i}\log \det\|g\| \ \ .
\eeqa
Then, with
\beq\label{D13}
F \ =\ \det\|K_{ij}\|
\eeq
(K being a 2Nx2N skew-symmetric matrix ), we find
\beq\label{D14}
V_{i}  =  {1\over 2}\partial_{i}\log \frac{(\det\|g\|)^{2}}{F} \hspace{2cm}
V_{\bar{i}}  =  {1\over 2}\partial_{\bar{i}}\log F
\eeq
As a consequence $FF^{\ast} \ \propto (\det\|g\|)^{2} $ . We emphasize that
{\sl a priori} F is a {\sl complex function}. This will be important in the
following (subsection 5.2). Notice also that $\Ga_{\mu}$ of equation
(\ref{b32}) is a true gradient vector :
$$ \Ga_{\mu} = -i\ \partial_{\mu}\log \frac{\det \|g\|}{F} $$ and then that
$C_{\mu\nu}$ vanishes. This is not a surprise as the holonomy is now Sp(N)
$\subset$ SU(2N).

The function F being rescaled such that
\beq\label{D15}
FF^{\ast} \ =\ (\det\|g\|)^{2},
\eeq
which has no consequence on the vector $V_{\mu}$, equations (\ref{b24}) give :
\beq\label{D16}
 Ric_{ij} = D_{j}V_{i} \ \ ,\ \ Ric_{i\,\bar{j}} = D_{\,\bar{j}}V_{i}
\eeq
We obtain the special case of ``generalised quasi Ricci flatness" ( see
(\ref{b28}) for $W = V$ and $\chi = -V$ ) that leads to SU(2N) holonomy (see
subsection (2.4)).

{\sl To summarize : (4,0) heterotic geometry implies, whatever the dimension of
the manifold be, generalised quasi Ricci flatness
\beq\label{D17}
Ric_{\mu\nu}\ =\  D_{\nu}V_{\mu}\
\eeq
where the vector $V_{\mu}$ , related to the torsion tensor, is given by}
\beq\label{D18}
V_{i} \  =  T^{k}_{ki}  = \ {1\over2}\partial_{i}\log F^{*} \hspace{1cm},\ \
V_{\bar{i}} \  =  T^{\bar{k}}_{\bar{k}\bar{i}}  = \
{1\over2}\partial_{\bar{i}}\log F
\hspace{1cm},\ \  FF^{\ast} \ =\ (\det\|g\|)^{2}.
\eeq

Notice that, in the absence of torsion, the Ricci tensor vanishes
\cite{[2]},\cite{[4]}. Moreover, equation (\ref{D17}) generalises in 4N
dimensions the corresponding one found in 4 dimensions \cite{[6]} where
$V_{\mu}$ is the dual of the torsion tensor
\beq\label{D19}
V_{\mu}^{4dim.} \ =\ {1\over 3!}\epsilon_{\mu\nu\rho\si}T^{\nu\rho\si}
\eeq

Finally, the scalar curvature reads
\beq\label{D20}
R\ =\ g^{i\bar{j}}(D_{i}V_{\bar{j}}\ +\ D_{\bar{j}}V_{i})\ =\  {1\over
2}g^{i\bar{j}}(D_{i}D_{\bar{j}}\log F \ +\ D_{\bar{j}}D_{i}\log F^{*})
\eeq
On the contrary of 4 dimensional case (as a consequence of equ. (\ref{c2})), it
no longer vanishes on general grounds.

If the function F is real (F = $\det\|g\|$), equation (\ref{D18}) shows that
$V_{\mu}$ is a gradient, which reminds us the conformal equivalence relation
(\ref{b264}). The conformal factor would then be $$ e^{-2f} =
(F)^{\frac{-1}{2(N-1)}}\ \ .$$ However, equation (\ref{b2622}), the other
condition for conformal invariance, does not hold on general grounds for N
$\neq$ 2. We then specify to 4 real dimensions manifolds where, due to
(\ref{c1}), it holds true. We are then in position to compare our results to
previous ones obtained by Callan, Harvey and Strominger (in the appendix of
\cite{[5]}). They claimed that there exists a conformal rescaling of the
metric,
\beq\label{D21}
\hat{g} \ =\ (\exp{2\phi})g
\eeq
where $\phi$ is defined through equations similar to (\ref{D13}) and the first
of (\ref{D121}) ( $\phi $ of ref. \cite{[5]} is equal to $-{1\over4}\log F$),
such that an equation similar to (\ref{c333}) holds :
\beq\label{D211}
\hat{\De}\Om \equiv \hat{\De}e^{-2\phi} = \hat{\De}e^{2f} = 0,
\hspace{1cm} \De \phi = \De f = 0 \ \ .
\eeq
Unfortunately, they implicitly suppose that their function $\phi$ is real,
which, as will be shown on an explicit example (equ. (\ref{e35}) in subsect.
5.2), is generally wrong. As a consequence and as explained in subsection
(3.2), the``metric" $\hat{g}$ introduced by Callan et al., does not lead to a
real distance, which is unacceptable. Then the assertion that``{\sl N=4  world
sheet supersymmetry imply  that the [coresponding] sigma model metric is
conformal to a Ricci flat K\"ahler metric} " is generally wrong. Rather, we
have shown that the metric is a generalised quasi Ricci flat one with a
vanishing scalar curvature.
Notice also that they forget about the second of  (\ref{D121}), then missing
the relationship between $\phi$ and $\det\|g\|$. Finally, the correct version
of equation (\ref{D211}) is not the Laplace condition on function $\phi$ when
$\phi$ is a true complex function, but, using equation (\ref{c2}), valid in two
complex dimensions, and result (\ref{D18}) :
\beqa\label{D22}
g^{i\bar{j}}D_{i}\partial_{\bar{j}}\phi \ & = & \
g^{i\bar{j}}D_{\bar{j}}\partial_{i}\phi^{*} \ =\ 0 \nnb\\
\Rightarrow \ \De\phi & \equiv & g^{i\bar{j}}(D_{i}\partial_{\bar{j}} +
D_{\bar{j}}\partial_{i})\phi\ =\ g^{i\bar{j}}D_{\bar{j}}\partial_{i}
(\phi-\phi^{\star}) \neq 0 \ \ .
\eeqa

We now construct some examples of (4,0) heterotic geometries in complex
coordinates.

$$ \ $$

\section{Examples of (4,4) and (4,0) geometries in complex coordinates}
In the construction of four dimensional hyperk\"ahler metrics, the most useful
tool was certainly the curvature self-duality requirement. This led to
Eguchi-Hanson \cite{[12]} and Taub-NUT metrics which are particular cases of
the multicentre metrics \cite{[11]}. More recently, but following the same
technique, Atiyah and Hitchin have obtained a genuinely new hyperk\"ahler
metric \cite{[14]} which is deeply related with the dynamics of a system of two
magnetic monopoles \cite{[15]}.

\noindent For all these metrics, at least one choice of holomorphic coordinates
is known. They are given in \cite{[16]} for Taub-NUT, in \cite{[17]} for
Eguchi-Hanson, in \cite{[16]} for the multicentre and in \cite{[19]} for
Atiyah-Hitchin.
Furthermore, there are important examples, mainly due to Calabi, for which the
use of holomorphic coordinates was essential \cite {[10]},\cite {[20]}.

Similarly, (4,0) geometries have been recently obtained by Delduc and Valent
\cite{[6]} as extensions of Taub-NUT and Eguchi-Hanson metrics with torsion,
using harmonic superspace and curvature self-duality. It is the aim of this
work to explore the advantages of holomorphic coordinates : we shall give
examples of these ones and compute for every case the vector $ V_{i}$ and the
function F of the preceding section.

\subsection {Eguchi-Hanson (with and without torsion)}
We extract from \cite{[6]} the vierbeins and the complex structures of
Eguchi-Hanson metric with torsion :
\beqa\label{e1}
e_{0}\ =\ {1\over2}\ga_{0}(s)[\sqrt{\frac{s}{s^{2}-a^{2}}}ds\ +\
\frac{2\rho}{\sqrt{s}}\eta_{3}] & ; & e_{3}\ =\
\ga_{0}(s)\sqrt{\frac{s^{2}-a^{2}}{s}}\eta_{3} \nnb\\
e_{1,2}\ =\ \ga_{0}(s)\sqrt{s}\eta_{1,2}  &{\rm  with } &
\frac{d}{ds}\ga_{0}^{2}(s)\ =\ \frac{L}{s^{2}-a^{2}+\rho^{2}}
\eeqa\
The one forms $\eta_{i}$ (i=1,2,3) satisfy
\beq\label{e2}
d\eta_{i}\ =\ -\epsilon_{ijk}\eta_{j}\wedge\eta_{k}
\eeq
and can be parametrised in sherical coordinates as :
\beqa\label{e3}
\eta_{1}\ & = & {1\over2}(\cos \phi\ d\theta\ +\ \sin \phi \sin \theta\ d\psi)
\nnb\\
\eta_{2}\ & = & {1\over2}(-\sin \phi\ d\theta\ +\ \cos \phi \sin \theta\ d\psi)
\nnb\\
\eta_{3}\ & = & {1\over2}(d\phi\ +\ \cos\theta\ d\psi)
\eeqa
The euclidean distance is $(x^{\mu} \equiv [s,\theta,\phi,\psi])$
\beq\label{e4}
d\tau^{2}\ =\ e_{0}^{2}\ +\sum_{i=1}^{3}e_{i}^{2} \ = \
g_{\mu\nu}dx^{\mu}dx^{\nu}
\eeq
and the complex structures two forms are
\beq\label{e5}
\omega_{i} \ =\ e_{0}\wedge e_{i}\ - {1\over2}\epsilon_{ijk}e_{j}\wedge e_{k}\
\ .
\eeq

One can check that in the following change of coordinates :
\beqa\label{e6}
x^{\mu}[s,\theta,\phi,\psi] & \rightarrow & y^{\mu}[s'=s,\theta ' =\theta,\phi
' =\phi + \beta\ -\phi_{0}(s),\psi ' =\psi]\nnb\\
{\rm with\ \ }  \tan \phi_{0}(s) \ & = & \frac{\rho}{\sqrt{s^{2}-a^{2}}}\ ,\
\beta \ \  {\rm   constant}\ ,
\eeqa
$ g_{\mu\nu}(\rho,a^{2},\ga_{0}(s)) $ transforms to $g'_{\mu\nu}\ \equiv
g_{\mu\nu}(\rho =0,a'^{2}=a^{2}-\rho^{2},\ga_{0}(s)) $. This means that under a
conformal rescaling $ \ga_{0}^{2}(s)$, Eguchi-Hanson metric with torsion
transforms to Eguchi-Hanson without torsion, in accordance with Callan et al.'s
claim \cite{[5]}. With regard to the complex structures
\beq\label{e7}
\Omega \ =\ \cos \alpha\ \omega_{3}\ + \sin \alpha\ (\cos \beta\ \omega_{1}\ +\
\sin \beta\ \omega_{2})\ \equiv\frac{1}{2}\Omega_{\mu\nu}dx^{\mu}\wedge
dx^{\nu}
\eeq
one obtains :
\beqa\label{e8}
\Omega_{s \theta} & = & \frac{\ga_{0}^{2}}{4}\frac{s}{\sqrt{s^{2}-a^{2}}}\sin
\alpha \cos (\beta + \phi) \nnb\\
\Omega_{s \phi} & = & \frac{\ga_{0}^{2}}{4}\cos \alpha \nnb\\
\Omega_{s \psi} & = & \frac{\ga_{0}^{2}}{4}\left[\cos \alpha \cos \theta
+\frac{s}{\sqrt{s^{2}-a^{2}}}\sin \alpha \sin \theta \sin (\beta + \phi)\right]
\nnb\\
\Omega_{\theta \phi} & = & \frac{\ga_{0}^{2}}{4}\sqrt{s^{2}-a^{2}+\rho^{2}}\sin
\alpha \sin  (\beta + \phi -\phi_{0}(s)) \nnb\\
\Omega_{\theta \psi} & = & \frac{\ga_{0}^{2}}{4}\left[-s\cos \alpha \sin \theta
\ +\sqrt{s^{2}-a^{2}+\rho^{2}}\sin \alpha \sin  (\beta + \phi
-\phi_{0}(s))\right] \nnb\\
\Omega_{\phi \psi} & = & \frac{\ga_{0}^{2}}{4}\sqrt{s^{2}-a^{2}+\rho^{2}}\sin
\alpha \cos  (\beta + \phi -\phi_{0}(s))
\eeqa
One can also check that in coordinates $y^{\mu}$

$$\Omega '_{\mu\nu}\ \equiv\ \Omega_{\mu\nu}(\rho
=0,a'^{2}=a^{2}-\rho^{2},\ga_{0}(s)) $$
As a consequence, the search for an holomorphic system of coordinates
associated to the complex structure
\beq\label{e9}
J \ =\ \cos \alpha\ J_{3}\ + \sin \alpha\ (\cos \beta\ J_{1}\ +\ \sin \beta\
J_{2})
\eeq
is the same as in the torsionless case.

In the looked-for coordinate system $ z^{\mu} (z^{i},\bar{z}^{i}) $, the
complex struture tensor $ \Omega_{\mu}^{\nu} $ must take the form (\ref{b8}) :
\beq\label{e10}
\Omega_{\mu}^{'j}\ = \ i \delta_{\mu}^{j}\ \ \ \, \ \Omega_{\mu}^{'\bar{j}}\ =
\ -i \delta_{\mu}^{\bar{j}}
\eeq
{}From $$ \Omega_{\mu}^{'\nu}\ =\ \frac{\partial x^{\rho}}{\partial
z^{\mu}}\Omega_{\rho}^{\sigma}\frac{\partial z^{\nu}}{\partial x^{\sigma}} $$
we obtain the conditions :
\beq\label{e11}
(\Omega_{\mu}^{\nu}\ -\ i\delta_{\mu}^{\nu})\partial_{\nu}z^{i}\ =\ 0\ \ \ ,\ \
i=1,2.
\eeq
This homogeneous system of four partial derivatives equations can be shown to
reduce itself to two independent equations  :
\beq\label{e12}
-i\partial_{s'}z\ +\ \frac{s\cos \alpha}{s^{2}-a'^{2}}\partial_{\phi '}z\ +\
\frac{\sin \alpha}{\sqrt{s^{2}-a'^{2}}}\left[\cos \phi '\partial_{\theta '}z\
+\ \frac{\sin \phi '}{\sin \theta '}(\partial_{\psi '}z\ -\ \cos \theta
'\partial_{\phi '}z)\right] \ =\ 0
\eeq
\beqa\label{e13}
i\partial_{\theta '}z\ & - & \frac{\cos \alpha}{\sin \theta '}(\partial_{\psi
'}z\ -\ \cos \theta '\partial_{\phi '})  + \nnb\\
& + & \sin \alpha \left[\sqrt{s^{2}-a'^{2}}\cos \phi '\partial_{s'}z\ -\
\frac{s}{\sqrt{s^{2}-a'^{2}}}\sin \phi '\partial_{\phi '}z\right] \ =\ 0
\eeqa
Special solutions are obtained for $\alpha\ =\ 0\ \  ( J_{3} $
diagonal)\cite{[17]} :
\beqa\label{e14}
z^{1} & = & (s^{2}-a'^{2})^{1/4}\cos (\theta '/2)\exp i(\phi '+\psi ')/2 \nnb\\
z^{2} & = & (s^{2}-a'^{2})^{1/4}\sin (\theta '/2)\exp i(\phi '-\psi ')/2
\eeqa
and for $\alpha\ =\ \pi/2\ \  ( J_{1} $ diagonal)
\beqa\label{e15}
z^{1} & = & \sqrt{s^{2}-a'^{2}}\sin \theta '\cos \phi ' \ -\ is\cos \theta '
\nnb\\
z^{2} & = & i\psi '\ +\ \log \frac{\sqrt{s^{2}-a'^{2}}(\cos \theta '\cos \phi '
\ +\ i\sin \phi ')\ +\ is\sin \theta '}{\sqrt{(z^{1})^{2}\ +\ a'^{2}}}
\eeqa
In the coordinate system (\ref{e14}), one obtains
\beqa\label{e16}
\omega_{1}\ -\ i\omega_{2} & = & (\ga_{0}(s))^{2}dz^{1}\wedge dz^{2} \nnb\\
\omega_{3} & = & ig_{i\,\bar{j}}dz^{i}\wedge d\bar{z}^{j}
\eeqa
with
\beqa\label{e17}
g_{1\bar{1}} & = &
\frac{(\ga_{0}(s))^{2}}{2}\left[\frac{s}{s^{2}-a'^{2}}|z^{2}|^{2}\ +\
\frac{1}{s}|z^{1}|^{2}\right] \nnb\\
g_{2\bar{2}} & = &
\frac{(\ga_{0}(s))^{2}}{2}\left[\frac{s}{s^{2}-a'^{2}}|z^{1}|^{2}\ +\
\frac{1}{s}|z^{2}|^{2}\right] \nnb\\
g_{1\bar{2}} & = &
-\frac{(\ga_{0}(s))^{2}}{2}\frac{a'^{2}}{s(s^{2}-a'^{2})}z^{2}z^{\bar{1}}\nnb\\
|z^{1}|^{2}\ +\ |z^{2}|^{2} & \equiv & t\ =\ \sqrt{s^{2}-a'^{2}}
\eeqa
and $$ \det\|g\| \ =\ \frac{(\ga_{0}(s))^{4}}{4}\ .  $$

The torsion tensor is then obtained as
\beq\label{e18}
T\ =\ \frac{L}{2(s^{2}-a'^{2})}[(\bar{z}^{2}d\bar{z}^{1}\ -\
\bar{z}^{1}d\bar{z}^{2})dz^{1}\wedge dz^{2}\ +\ c.c.]
\eeq
and its closedness is readily verified.

Notice that with $\ga_{0}^{2}$ changed to $\ga\ ,\ t\ =\  \sqrt{s^{2}-a'^{2}}$
changed to s and $a'^{2}$ to $-\la^{2}$, one reproduces equations
(\ref{c11},\ref{c144}).

The vector $V_i$ and function F are then obtained from equations
(\ref{D11},\ref{D13}, \ref{D14} and \ref{e16}) :
\beq\label{e19}
V_i\ =\ {1\over 2}\partial_i \log (\frac{(\ga_{0}(s))^{2}}{2})^2 \ \ ; \ \ F\
=\ F^*\ =\ \left(\frac{(\ga_{0}(s))^{2}}{2}\right)^2
\eeq
and
\beqa\label{ee5}
d\omega _3 & = & \frac 12\omega _3\wedge d(\log \det ||g||)\\ \label{ee6}
d(\omega _1-i\omega _2) & = & \frac 12(\omega _1-i\omega _2)\wedge d(\log F).
\eeqa
Here it is obvious that Callan et al' s assertion works : indeed Eguchi-Hanson
with torsion is conformally equivalent to its torsionless counterpart, with the
conformal factor
$$e^{2f}=\frac {(\gamma_0(s))^2}{2}.$$

The usual hyperk\"ahler Eguchi-Hanson metric is obtained for L = 0 ( $\ga_{0} $
constant which may be taken to be 1) and, still in coordinates (\ref{e14}) the
K\"ahler potential is K(t) with :
\beq\label{20}
\frac {dK}{dt}\ =\ \frac{\sqrt{t^2 + a^2}}{2t}
\eeq
In coordinates (\ref{e15}), corresponding to $J_1$ diagonal, we found in the
same limit L = $\rho$ = 0
\beq\label{e21}
K\ =\ -\frac{s}{2}.
\eeq

\subsection{Taub-NUT (with and without torsion)}
Here too we extract from \cite{[6]} the vierbeins and the complex structures :
\beqa\label{e22}
e_{0}\ =\ {1\over2}\ga_{0}(s)a(s)[\frac{1+\la s}{\sqrt{s}}ds\ +\ 2\rho
s^{3/2}\si_{3}] & ; & e_{3}\ =\ \ga_{0}(s)a(s)\sqrt{s}\si_{3} \nnb\\
e_{1,2}\ =\ \ga_{0}(s)a(s)(1+\la s)\sqrt{s}\si_{1,2}  &{\rm with } & a(s)\ =\
{1\over{\sqrt{(1+\la s)(1+\rho^2s^2)}}} \nnb\\
 & {\rm and } & \frac{d}{ds}\ga_{0}^{2}(s)\ =\ \frac{L}{s^{2}}
\eeqa\
The one forms $\si_{i}$ (i=1,2,3) satisfy
\beq\label{e23}
d\si_{i}\ =\ \epsilon_{ijk}\si_{j}\wedge\si_{k}
\eeq
and can be parametrised in sherical coordinates as :
\beqa\label{e24}
\si_{1}\ & = & {1\over2}(-\sin \psi\  d\theta\ -\ \cos \psi \sin \theta\
d\phi) \nnb\\
\si_{2}\ & = & {1\over2}(\cos \psi\  d\theta\ -\ \sin \psi \sin \theta\  d\phi)
\nnb\\
\si_{3}\ & = & {1\over2}(d\psi\ -\ \cos\theta\  d\phi)
\eeqa
The euclidean distance is $(x^{\mu} \equiv [s,\theta,\phi,\psi])$
\beq\label{e25}
d\tau^{2}\ =\ e_{0}^{2}\ +\sum_{i=1}^{3}e_{i}^{2} \ = \
g_{\mu\nu}dx^{\mu}dx^{\nu}
\eeq
Let us define
\beq\label{e26}
{\cal K}_{i} \ =\ e_{0}\wedge e_{i}\ - {1\over2}\epsilon_{ijk}e_{j}\wedge
e_{k}\ \ ;\ \ \ i,j,k\ =\ 1,2,3
\eeq
from which we deduce the complex structures two forms
\beqa\label{e28}
\omega_{1} \ \pm i\omega_{2} & = & [(\pm i\sin \psi - \cos \theta \cos
\psi){\cal K}_1 \ -\ (\pm i\cos \psi + \cos \theta \sin \psi ){\cal K}_2 \
\nnb\\
& &\hspace{8cm}+\ \sin \theta \ {\cal K}_3]\exp{ \pm i\phi} \nnb\\
\omega_{3} & = & \sin \theta[\cos \psi \ {\cal K}_1 \ + \ \sin \psi \ {\cal
K}_2 ]\ +\ \cos \theta \ {\cal K}_3.
\eeqa

We now look for a new system of coordinates $ z^1\ ,z^2\ ,\bar{z}^{1}\ ,
\bar{z}^{2} $ wich diagonalize  the complex structures according to (\ref{b8}).
With $\Omega$ defined through (\ref{e7}), one obtains
\beqa\label{e29}
\Omega_{s \theta} & = & \frac{\ga_{0}^{2}}{4(1+\rho^2 s^2)}(1+\la s)\sin \alpha
\sin (\phi - \beta) \nnb\\
\Omega_{s \phi} & = & \frac{\ga_{0}^{2}}{4(1+\rho^2 s^2)}[\la s\sin \al \sin
\theta \cos \theta \cos (\phi - \beta)\ -\ \cos \alpha\ (1+\la s \sin^2\theta)]
\nnb\\
\Omega_{s \psi} & = & \frac{\ga_{0}^{2}}{4(1+\rho^2 s^2)}[\sin \alpha \sin
\theta \cos (\phi - \beta)\ +\ \cos \alpha \cos \theta\ ] \nnb\\
\Omega_{\theta \phi} & = & -\frac{s\ga_{0}^{2}}{4(1+\rho^2 s^2)}[\sin \alpha
[(1+\la s \sin^2\theta)\cos(\phi - \beta)\ -\ \rho s\cos \theta \sin (\phi -
\beta)]\ +\nnb\\
& & \hspace{10 cm}+\  \la s\cos \al \sin \theta \cos \theta] \nnb\\
\Omega_{\theta \psi} & = & \frac{s\ga_{0}^{2}}{4(1+\rho^2 s^2)}[\sin \alpha
[\cos \theta \cos(\phi - \beta)\ -\ \rho s \sin  (\phi - \beta)]\ -\ \cos
\alpha \sin \theta] \nnb\\
\Omega_{\phi \psi} & = & -\frac{s\ga_{0}^{2}\sin \theta}{4(1+\rho^2 s^2)}[\sin
\alpha [\sin (\phi - \beta)\ +\ \rho s\cos \theta \cos (\phi - \beta)]\ -\ \rho
s\cos \al \sin \theta]
\eeqa
Here again, equations (\ref{e11}) give an homogeneous system of four partial
derivatives equations which can be shown to reduce itself to two independent
ones. For brevity, we give them only in the case $\alpha\ =\ 0\ \  ( J_{3} $
diagonal) :
\beq\label{e30}
(i\ +\ \rho s \cos \theta)s\partial_{s}z\ +\ \partial_{\phi}z\ -\ \la s \cos
\theta\ \partial_{\psi }z\  =\ 0
\eeq
\beq\label{e31}
\sin^2 \theta\ s\partial_{s}z\ +\ \sin \theta \cos \theta\ \partial_{\theta}z\
-\ i(\partial_{\phi}z\ +\ \cos \theta\ \partial_{\psi}z)\ =\ 0
\eeq
Special solutions may be obtained in the form :
\beqa\label{e32}
z^{1} & = & -i\phi\ +\ \log (s \sin \theta)\ -\ \log (1 - i\rho s \cos \theta)
\nnb\\
z^{2} & = & i\psi -\ \log (\tan {\theta\over 2})\ +\ \frac{i\la}{\rho}\log (1 -
i\rho s \cos \theta)
\eeqa
In this coordinate system, one obtains
\beqa\label{e33}
\omega_{1}\ -\ i\omega_{2} & = & -i\frac{s(\ga_{0}(s))^{2}}{4}\frac{(1 - i\rho
s \cos \theta)\sin \theta}{(1 + \rho^2 s^2)}\exp {-i\phi}\ \ dz^{1}\wedge
dz^{2} \nnb\\
\omega_{3} & = & ig_{i\,\bar{j}}dz^{i}\wedge d\bar{z}^{j}
\eeqa
with
\beqa\label{e34}
g_{1\bar{1}} & = & \frac{s(\ga_{0}(s))^{2}}{8}\left[\frac{\cos^2 \theta}{1 +
\la s}\ +\ \frac{1+ \la s}{1 + \rho^2 s^2}\sin^2 \theta\right] \nnb\\
g_{2\bar{2}} & = & \frac{s(\ga_{0}(s))^{2}}{8}\frac{1}{1 + \la s} \nnb\\
g_{1\bar{2}} & = & \frac{s(\ga_{0}(s))^{2}}{8}\left[\frac{\cos \theta}{1 + \la
s}\ +\ i\ \frac{\rho s}{1 + \rho^2 s^2}\sin^2 \theta\right]
\eeqa
and $$ \det\|g\| \ =\ \left[\frac{s(\ga_{0}(s))^{2}}{8}\right]^2 \frac{(1 +
\rho^2 s^2 \cos^2 \theta)\sin^2 \theta}{(1 + \rho^2 s^2)^2} \ . $$

The torsion tensor is then obtained but, as its expression is lenghty, we do
not give it here. The vector $V_i$ and function F then result from equations
(\ref{D11},\ref{D13},\ref{D14} and \ref{e33}) :
\beq\label{e35}
V_i\ =\ {1\over 2}\partial_i \log F^{*} \ \ \ ; \ F\ =\ \
\left[\frac{s(\ga_{0}(s))^{2}}{8}\right]^2 \frac{(1 - i\rho s \cos
\theta)^2\sin^2 \theta}{(1 + \rho^2 s^2)^2}\exp {-2i\phi}
\eeq
and we have
\beqa
d\omega_3 & = & \frac 12\omega_3\wedge\left(d(\log{\rm
det}||g||)+\frac{i}{2}d^c\log(\frac{F}{F^*})\right)\nnb\\
d(\omega_1-i\omega_2) & = & \frac 12(\omega_1-i\omega_2)\wedge d\log F \nnb\\
d(\omega_1+i\omega_2) & = & \frac 12(\omega_1+i\omega_2)\wedge
d\log F^*,\nnb
\eeqa
with the notations
%% FOLLOWING LINE CANNOT BE BROKEN BEFORE 80 CHAR
%% FOLLOWING LINE CANNOT BE BROKEN BEFORE 80 CHAR
$$d=d'+d'',\hspace{0.5cm}d^c=i(d'-d''),\hspace{0.5cm}d'f=z^i\frac{\partial}{\partial z^i}f,\hspace{0.5cm}d''f={\overline{z}^i}\frac{\partial}{\partial\overline {z}^i}f.$$

Notice the important difference with Eguchi-Hanson : here {\sl the function F
is  a complex one}. One should wonder whether, in an analytic change of
coordinates, the function F might become real. F being a determinant, $ \log F
\  \rightarrow \ \  \log F + h(z^i) \ $. Then
$$\log \frac{F}{F^{*}} \  \rightarrow \ \ \log \frac{F'}{F'^{*}} = \log
\frac{F}{F^{*}} + h(z^i) - \bar{h}(\bar{z}^i)\  .$$ A real F' then needs
$$\partial_i \partial_{\bar{j}}\left[ \log \frac{F}{F^{*}} \right]
\equiv 0 \ .$$
 One computes  $$\log \frac{F}{F^{*}} = -4i \phi + 2\log \frac{(1 - i\rho s
\cos \theta)}{((1 + i\rho s \cos \theta)} = -2(z^1 - \bar{z}^1) - 8i\phi$$
and, for example  $$\partial_2 \partial_{\bar{2}}\left[ \log \frac{F}{F^{*}}
\right] = -8i \partial_2 \partial_{\bar{2}}\phi = -8i\partial_2\left[\frac{\rho
s}{2(1+\la s)}\right] = -2i\rho s\cos \theta \frac{1+ \rho^2 s^2}{(1+ \la s)^3}
\neq 0\ \ .  $$ As a consequence, there is no conformal equivalence between
this (4,0) geometry and an hyperk\"ahler one and this gives a counterexample to
Callan et al.'s assertion.

Let us observe that, contrarily to Eguchi-Hanson's case, the new parameter
$\rho$ plays a distinguished role as it signals a new geometry. The second
parameter in the torsion, $L$ ($\Leftrightarrow$ the function $\gamma_0^2(s)$),
 can be re-absorbed by a conformal transformation which, by the way, is missing
in the present status of the Harmonic superspace approach \cite{[6]}.

The usual hyperk\"ahler Taub-NUT metric is obtained for $\rho$ = L = 0
( $\ga_{0}(s) $ constant which may be taken to be 1) and, in coordinates
(\ref{e32}) the K\"ahler potential is :
\beq\label{e36}\displaystyle
K\ =\ \frac {s}{2}\left[1 + \frac 14 \la s(1+ \cos^2\theta)\right].
\eeq

\section{Concluding remarks}
We think to have clearly exemplified the advantages of using holomorphic
coordinates to describe heterotic geometry. This enables us to obtain new and
easy derivations of Eguchi-Hanson metric with torsion \cite{[6]} and
generalisations of the K\"ahler quasi Ricci flat metrics of Bonneau and Delduc
\cite{[55]} and of the scalar flat metrics of LeBrun \cite{[44]}.

We have also proven that, contrarily to Callan et al.'s claim \cite{[5]}, even
for a 4 dimensional manifold, (4,0) heterotic geometry is not simply
conformally equivalent to a (4,4) HyperK\"ahler one.

\noindent It may be interesting to summarise the results in the language of the
holonomy group :

\begin{center}
\begin{tabular}{|l||l|l|}
\hline
\multicolumn{1}{|c||}{}
& \multicolumn{1}{c|}{}
& \multicolumn{1}{c|}{}\\
\ \ \ {PROPERTIES}\ \ \ &\ \ \ \ \ \ \ \ \ {no torsion}\ &\ \ \ \ \ \ \ \ \
{with torsion}\ \\
  &  &  \\ \hline \hline  &  &  \\
real dimension N  & susy (1, 1) & (1, 0) heterotic geometry\\
 holonomy O(N) &\phantom{(1, 0) heterotic geometry\ \ \  }  &
\phantom{(1, 0) heterotic geometry\ \ \  }   \\ &  &  \\ \hline
 &  &  \\

real dimension 2N   & susy (2, 2)  & (2, 0) heterotic geometry\\
holonomy U(N) & K\"{a}hler manifold &  \\  &  &  \\
\hline &  &  \\

real dimension 2N  & susy (2, 2)  &  (2, 0) heterotic geometry\\
holonomy SU(N) & K\"{a}hler + Ricci flat & special quasi Ricci flat space \\
  &  & (W=-$\chi$=V where V is a vector  \\
  &   & related to the torsion tensor) \\  &  &  \\
\hline
 &  &  \\
real dimension 4N\ \ \  & susy (4, 4)  & (4, 0) heterotic geometry \\
holonomy Sp(N) & hyperk\"{a}hler ($\rightarrow$ Ricci flat)\ \ \ \ \ \  &
special quasi Ricci flat space\ \ \ \ \ \  \\
  &  & (W=-$\chi$=V where V depends on \\
  &   & a single complex function F)  \\  &  &  \\
 \hline
\end{tabular}
\end{center}
$$  $$
\noindent{\bf Acknowledgements :} the authors are happy to thank F. Delduc for
useful discussions.

\bibliographystyle{plain}

\end{document}